\documentclass[aps,preprint]{revtex4}
\usepackage{makeidx}
\usepackage{amsmath}
\usepackage{graphicx}
\usepackage{epstopdf}

\begin{document}

\title{Surface Terms of Quartic Quasitopological Gravity and
Thermodynamics of Nonlinear Charged Rotating Black Branes}
\author{A. Bazrafshan $^{1}$, M. H. Dehghani $^{2,3}$ \footnote{%
email address: mhd@shirazu.ac.ir}, and M. Ghanaatian $^{4} $}
\affiliation{$^1$ Department of Physics, Jahrom University, 74137-66171 Jahrom, Iran}
\affiliation{$^2$ Center for Excellence in Astronomy \& Astrophysics (CEAA -- RIAAM) --
Maragha, IRAN, P. O. Box: 55134 - 441\\
$^3$Physics Department and Biruni Observatory, College of Sciences, Shiraz
University, Shiraz 71454, Iran}
\affiliation{$^4$ Department of Physics, Payame Noor University, Iran}

\begin{abstract}
As in the case of Einstein or Lovelock gravity, the action of quartic
quasitopological gravity has not a well-defined variational principle. In
this paper, we first introduce a surface term that makes the variation of
quartic quasitopological gravity well defined. Second, we present the static
charged solutions of quartic quasitopological gravity in the presence of a
non linear electromagnetic field. One of the branch of these solutions
presents a black brane with one or two horizons or a naked singularity
depending on the charge and mass of the solution. The thermodynamic of these
black branes are investigated through the use of the Gibbs free energy. In
order to do this, we calculate the finite action by use of the counterterm
method inspired by AdS/CFT correspondence. Introducing a Smarr-type formula,
we also show that the conserved and thermodynamics quantities of these
solutions satisfy the first law of thermodynamics. Finally, we present the
charged rotating black branes in $(n+1)$ dimensions with $k\leq [n/2]$
rotation parameters and investigate their thermodynamics.
\end{abstract}

\maketitle


\section{Introduction}

The anti-de Sitter/conformal field theory (AdS/CFT) correspondence \cite%
{AdS/CFT} provides a prescription to compute vacuum expectation values of
the $n$-dimensional conformal field theory (CFT) operators in terms of dual
classical field in an $(n+1)$-dimensional anti-de Sitter (AdS) spacetime.
This prescription has been checked successfully in many examples. It is now
known that the central charges in the $n$-dimensional conformal field theory
relate to the coupling constants of the $(n+1)$-dimensional gravity. But,
since Einstein gravity has only one coupling constant, it is only dual to
those conformal field theories for which all the central charges are equal.
In order to extend the duality to new classes of conformal field theories
with different ratios between central charges, one should consider theories
of gravity with enough parameters to account for the ratios between central
charges. In this case, one should add higher curvature terms with more
coupling constants within a perturbative framework to action \cite{Buch}.
The computation of central charges and their relation with the coupling
constants of Lovelock gravity \cite{Love} have been investigated for the
case of second and third order Lovelock gravity \cite{Myers1,Hofman}.
Lovelock gravity is the most general theory of gravitation which is
quasilinear in the second derivatives of the metric and does not contain any
higher derivatives in any dimension for a general spacetime. Although the
equations of motion of $n$th order Lovelock gravity are second-order
differential equations, the $n$th order Lovelock term has no contribution to
the field equations in $2n$ and lower dimensions and therefore it is useless
for studying field theories in four dimensions. In order to have
gravitational theory in five dimension with more coupling constants, one
should study higher derivative gravity or quasitopological gravity. Although
the field equations of higher derivative gravity contain higher derivatives
of the metric more than two, recently, third \cite{MyersRobin} and fourth
\cite{DBM} order quasitopological terms have been introduced which have
contribution to the field equations in five and higher dimensions, and the
equations of motion of these theories are second-order differential
equations when the metric is spherically symmetry. As in the case of third
and fourth order Lovelock gravities, the cubic and quartic terms have no
contribution to the field equations in six and eight dimensions,
respectively. Although this property of Lovelock gravity is due to a
topological origin of Euler density, this property of the new theory does
not have a topological origin and therefore these theories are known as
cubic and quartic quasitopological gravity. The black hole solutions and
holography study of cubic quasitopological gravities have been investigated
in Refs. \cite{Myers2,QTG,BDM}.

The main aim of this paper is to introduce a surface term for quartic
quasitopological gravity introduced in \cite{DBM} in order to have a
well-defined variational principle in the case of spacetimes with flat
boundary. As for the case of the Einstein-Hilbert action which does not have
a well-defined variational principle \cite{Gib}, the variation of the action
of quasitopological gravity is not well-defined. This is due to the fact
that one encounters a total derivative that produces a surface integral
involving the derivative of the variation of the metric normal to the
boundary. To cancel this normal derivative term, one has to add a surface
term to the action. The surface terms of Lovelock and cubic quasitopological
gravity have been introduced in \cite{SurLove,DBS} and \cite{DV},
respectively. Here, we introduce the surface term for quartic
quasitopological gravity. Another motivation of introducing the surfaceterm
is due to the fact that in the Hamiltonian formalism of general relativity,
the surface terms should be known \cite{York}.

Our second goal is to introduce the black brane solutions of quartic
quasitopological gravity in the presence of a nonlinear Maxwell field.
Having nonlinear terms of curvature in the action of gravity, it is natural
to assume nonlinear terms of electromagnetic tensor field, $(-F_{\mu \nu
}F^{\mu \nu })^{s}$, in the matter action \cite{Martinez}. Our third aim is
to introduce a counterterm that removes the divergences of the action and
conserved quantities of the black brane solutions. Finally, we investigate
the thermodynamics of static and rotating black brane solutions of quartic
quasitopological gravity in the presence of power law Maxwell field.

The outline of our paper is as follows. In Sec. \ref{Bound}, we give a brief
review of the quartic quasitopological gravity and review the surface terms
which make the variation of Einstein, Gauss-Bonnet and cubic
quasitopological terms well defined for flat boundary. Then we introduce the
surface term of quasitopological gravity for spacetimes with flat boundary.
In Sec. \ref{Solutions}, we introduce asymptotically AdS charged black
branes of quasitopological gravity in the presence of nonlinear Maxwell
field. Section \ref{Therm} is devoted to the investigation of the
thermodynamic properties of these solutions by using the relation between
the on shell action and Gibbs free energy. In Sec. \ref{Rot}, we endow our
solutions with rotation and study the thermodynamic properties of these
rotating charged black branes. Finally, we finish our paper with some
concluding remarks.

\section{\ Surface Terms of Quartic Quasitopological Gravity \label{Bound}}

The field equations of fourth order Lovelock gravity are at most second
order partial differential equations, while it contains fourth order terms
in Riemann tensor. But, the fourth order term has contribution in the field
equations provided one consider the field equations in nine and higher
dimensions. On the other hand, quartic quasitopological gravity is a theory
of gravity which contains fourth order terms in Riemann tensor with at most
second order differential equation, while the quartic term contributes to
the field equations in five and higher dimensions \cite{DBM}. Of course, one
should note that the field equations of quartic quasitopological gravity are
second order differential equations only for a spherically symmetric
spacetime, and has no contribution in eight dimensions. The action of fourth
order quasitopological gravity in $(n+1)$ dimensions in the presence of a
nonlinear electromagnetic field may be written as
\begin{equation}
I_{G}=\frac{1}{16\pi }\int d^{n+1}x\sqrt{-g}[-2\Lambda +\mathcal{L}_{1}+\mu
_{2}\mathcal{L}_{2}+\mu _{3}\mathcal{X}_{3}+\mu _{4}\mathcal{X}_{4}+L(F)],
\label{Act1}
\end{equation}%
where $\Lambda =-n(n-1)/2l^{2}$ is the cosmological constant of AdS
spacetime, $\mathcal{L}_{1}={R}$ is just the Einstein-Hilbert Lagrangian, $%
\mathcal{L}_{2}=R_{abcd}{R}^{abcd}-4{R}_{ab}{R}^{ab}+{R}^{2}$ is the second
order Lovelock (Gauss-Bonnet) Lagrangian, $\mathcal{X}_{3}$\ is the
curvature-cubed Lagrangian \cite{Myers1}
\begin{eqnarray}
\mathcal{X}_{3} &=&R_{ab}^{cd}R_{cd}^{\,\,e\,\,\,f}R_{e\,\,f}^{\,\,a\,\,\,b}+%
\frac{1}{(2n-1)(n-3)}\left( \frac{3(3n-5)}{8}R_{abcd}R^{abcd}R\right.  \notag
\\
&&-3(n-1)R_{abcd}R^{abc}{}_{e}R^{de}+3(n+1)R_{abcd}R^{ac}R^{bd}  \notag \\
&&\left. +\,6(n-1)R_{a}{}^{b}R_{b}{}^{c}R_{c}{}^{a}-\frac{3(3n-1)}{2}%
R_{a}^{\,\,b}R_{b}^{\,\,a}R+\frac{3(n+1)}{8}R^{3}\right) ,  \label{X3}
\end{eqnarray}%
and $\mathcal{X}_{4}$ is the fourth order term of quasitopological gravity
\cite{DBM}
\begin{eqnarray}
\mathcal{X}_{4}\hspace{-0.2cm} &=&\hspace{-0.2cm}c_{1}R_{abcd}R^{cdef}R_{%
\phantom{hg}{ef}%
}^{hg}R_{hg}{}^{ab}+c_{2}R_{abcd}R^{abcd}R_{ef}R^{ef}+c_{3}RR_{ab}R^{ac}R_{c}{}^{b}+c_{4}(R_{abcd}R^{abcd})^{2}
\notag \\
&&\hspace{-0.1cm}%
+c_{5}R_{ab}R^{ac}R_{cd}R^{db}+c_{6}RR_{abcd}R^{ac}R^{db}+c_{7}R_{abcd}R^{ac}R^{be}R_{%
\phantom{d}{e}}^{d}+c_{8}R_{abcd}R^{acef}R_{\phantom{b}{e}}^{b}R_{%
\phantom{d}{f}}^{d}  \notag \\
&&\hspace{-0.1cm}%
+c_{9}R_{abcd}R^{ac}R_{ef}R^{bedf}+c_{10}R^{4}+c_{11}R^{2}R_{abcd}R^{abcd}+c_{12}R^{2}R_{ab}R^{ab}
\notag \\
&&\hspace{-0.1cm}%
+c_{13}R_{abcd}R^{abef}R_{ef}{}_{g}^{c}R^{dg}+c_{14}R_{abcd}R^{aecf}R_{gehf}R^{gbhd},
\label{X4}
\end{eqnarray}%
with
\begin{eqnarray*}
c_{1} &=&-\left( n-1\right) \left( {n}^{7}-3\,{n}^{6}-29\,{n}^{5}+170\,{n}%
^{4}-349\,{n}^{3}+348\,{n}^{2}-180\,n+36\right) , \\
c_{2} &=&-4\,\left( n-3\right) \left( 2\,{n}^{6}-20\,{n}^{5}+65\,{n}^{4}-81\,%
{n}^{3}+13\,{n}^{2}+45\,n-18\right) , \\
c_{3} &=&-64\,\left( n-1\right) \left( 3\,{n}^{2}-8\,n+3\right) \left( {n}%
^{2}-3\,n+3\right) , \\
c_{4} &=&-{(n}^{8}-6\,{n}^{7}+12\,{n}^{6}-22\,{n}^{5}+114\,{n}^{4}-345\,{n}%
^{3}+468\,{n}^{2}-270\,n+54), \\
c_{5} &=&16\,\left( n-1\right) \left( 10\,{n}^{4}-51\,{n}^{3}+93\,{n}%
^{2}-72\,n+18\right) , \\
c_{6} &=&--32\,\left( n-1\right) ^{2}\left( n-3\right) ^{2}\left( 3\,{n}%
^{2}-8\,n+3\right) , \\
c_{7} &=&64\,\left( n-2\right) \left( n-1\right) ^{2}\left( 4\,{n}^{3}-18\,{n%
}^{2}+27\,n-9\right) , \\
c_{8} &=&-96\,\left( n-1\right) \left( n-2\right) \left( 2\,{n}^{4}-7\,{n}%
^{3}+4\,{n}^{2}+6\,n-3\right) , \\
c_{9} &=&16\left( n-1\right) ^{3}\left( 2\,{n}^{4}-26\,{n}^{3}+93\,{n}%
^{2}-117\,n+36\right) , \\
c_{10} &=&{n}^{5}-31\,{n}^{4}+168\,{n}^{3}-360\,{n}^{2}+330\,n-90, \\
c_{11} &=&2\,(6\,{n}^{6}-67\,{n}^{5}+311\,{n}^{4}-742\,{n}^{3}+936\,{n}%
^{2}-576\,n+126), \\
c_{12} &=&8\,{(}7\,{n}^{5}-47\,{n}^{4}+121\,{n}^{3}-141\,{n}^{2}+63\,n-9), \\
c_{13} &=&16\,n\left( n-1\right) \left( n-2\right) \left( n-3\right) \left(
3\,{n}^{2}-8\,n+3\right) , \\
c_{14} &=&8\,\left( n-1\right) \left( {n}^{7}-4\,{n}^{6}-15\,{n}^{5}+122\,{n}%
^{4}-287\,{n}^{3}+297\,{n}^{2}-126\,n+18\right).
\end{eqnarray*}%
In the action (\ref{Act1}), $L(F)$ is the Lagrangian of power Maxwell
invariant theory \cite{Martinez}
\begin{equation}
L(F)={(-F)}^{s}  \label{Lmat}
\end{equation}%
where $F=F_{\mu \nu }F^{\mu \nu }$, $F_{\mu \nu }=\partial _{\mu }A_{\nu
}-\partial _{\nu }A_{\mu }$ is the electromagnetic field tensor and $A_{\mu
} $ is the vector potential. One may note that in the limit $s=1$, $L(F)$
reduces to the standard Maxwell Lagrangian. Also, it is worth to mention
that this Lagrangian is trivial for $s=0$.

In general, the variation of the action (\ref{Act1}) with respect to the
metric does not have a well-defined variational principle in the case of a
manifold with boundary. This problem has been first noted gravity by Gibbons
and Hawking for Einstein. They suggested that the following boundary term
\cite{Gib}
\begin{equation}
I_{b}^{(1)}=\frac{1}{8\pi }\int_{\partial \mathcal{M}}d^{n}x\sqrt{-\gamma }K
\label{Ib1}
\end{equation}%
makes the Einstein-Hilbert action well defined, where $\gamma _{ab}$ and $K$
are the induced metric and the trace of extrinsic curvature of the boundary $%
\partial \mathcal{M}$. This problem has been investigated for Lovelock
gravity, and the boundary terms has been introduced in \cite{SurLove,DBS}.
Here, we are interested in the spacetimes with flat boundaries, $\hat{R}%
_{abcd}(\gamma )=0$, and therefore the proper surface term for Gauss-Bonnet
term is
\begin{equation}
I_{b}^{(2)}=\frac{1}{8\pi }\int_{\partial \mathcal{M}}d^{n}x\sqrt{-\gamma }%
\left\{ \frac{2\hat{\mu _{2}}l^{2}}{(n-2)(n-3)}J\right\} ,  \label{Ib2}
\end{equation}%
where $J$ is the trace of
\begin{equation}
J_{ab}=\frac{1}{3}%
(2KK_{ac}K_{b}^{c}+K_{cd}K^{cd}K_{ab}-2K_{ac}K^{cd}K_{db}-K^{2}K_{ab}).
\label{Jab}
\end{equation}%
The surface terms for the curvature-cubed term of quasitopological gravity
have been introduced in \cite{DV} as
\begin{eqnarray}
&&I_{b}^{(3)}=\frac{1}{8\pi }\int_{\partial \mathcal{M}}d^{n}x\sqrt{-\gamma }%
\Big\{\frac{3\hat{\mu _{3}}l^{4}}{5n(n-2)(n-1)^{2}(n-5)}%
(nK^{5}-2K^{3}K_{ab}K^{ab}  \notag \\
&&\,\ \ \ \ \ \ \ \ \ \ \ \ \ \ \ \ \ \ \ \ \ \ \ \ \ \ \ \
+4(n-1)K_{ab}K^{ab}K_{cd}K_{e}^{d}K^{ec}-  \notag \\
&&\,\ \ \ \ \ \ \ \ \ \ \ \ \ \ \ \ \ \ \ \ \ \ \ \ \ \ \ \
(5n-6)KK_{ab}[nK^{ab}K^{cd}K_{cd}-(n-1)K^{ac}K^{bd}K_{cd}])\Big\}.
\label{Ib3}
\end{eqnarray}

Here, we are going to present a surface term that makes the action of
quartic quasitopological gravity well defined. In order to do this, one
should combine different seventh power of extrinsic curvature:
\begin{eqnarray}
&&I_{b}^{(4)}={\frac{1}{8\pi }}\int_{\partial \mathcal{M}}d^{n}x\sqrt{%
-\gamma }{\frac{\hat{2\mu _{4}}{l}^{6}}{7n(n-1)\left( n-2\right) \left(
n-7\right) \left( {n}^{2}-3\,n+3\right) }}\Big\{\alpha
_{1}K^{3}K^{ab}K_{ac}K_{bd}K^{cd}  \notag \\
&&\,\ \ \ \ \ \ \ \ \ \ \ \ \ \ \ \ \ \ \ \ \ \ \ \ \ \ \ \ +\alpha
_{2}K^{2}K^{ab}K_{ab}K^{cd}K_{c}^{e}K_{de}+\alpha
_{3}K^{2}K^{ab}K_{ac}K_{bd}K^{ce}K_{e}^{d}  \notag \\
&&\,\ \ \ \ \ \ \ \ \ \ \ \ \ \ \ \ \ \ \ \ \ \ \ \ \ \ \ \ +\alpha
_{4}KK^{ab}K_{ab}K^{cd}K_{c}^{e}K_{d}^{f}K_{ef}+\alpha
_{5}KK^{ab}K_{a}^{c}K_{bc}K^{de}K_{d}^{f}K_{ef}  \notag \\
&&\,\ \ \ \ \ \ \ \ \ \ \ \ \ \ \ \ \ \ \ \ \ \ \ \ \ \ \ \ +\alpha
_{6}KK^{ab}K_{ac}K_{bd}K^{ce}K^{df}K_{ef}+\alpha
_{7}K^{ab}K_{a}^{c}K_{bc}K^{de}K_{df}K_{eg}K^{fg}\Big\}  \label{Ib4}
\end{eqnarray}%
Using the $(n+1)$-dimensional static metric with a flat boundary%
\begin{equation}
ds^{2}=-g(\rho )dt^{2}+\frac{d\rho ^{2}}{f(\rho )}+\rho
^{2}\;\sum_{i=1}^{n-1}d\phi _{i}{}^{2}.  \label{met1}
\end{equation}%
and the action (\ref{Ib4}), one may fix the unknown coefficients $\alpha
_{i} $'s such that the variation of $I_{b}^{(4)}$ removes all the normal
derivatives of\ $\delta g_{\mu \nu }$ of quartic quasitopological term. One
obtains
\begin{eqnarray}
\alpha _{1} &=&\,n^{2}(n-1),  \notag \\
\alpha _{2} &=&2\left( {n}^{3}-4\,{n}^{2}+3\,n+3\right),  \notag \\
\alpha _{3} &=&-6n(n-1)^{2},  \notag \\
\alpha _{4} &=&-4n\left( {n}^{2}-3\,n+3\right) (n+2),  \notag \\
\alpha _{5} &=&-\,2n\left( n-3\right) \left( {n}^{2}-n-3\right),  \notag \\
\alpha _{6} &=&2(n-1)\left( 3\,{n}^{3}-{n}^{2}-9\,n+12\right),  \notag \\
\alpha _{7} &=&24.  \label{alph}
\end{eqnarray}%
Thus, the boundary term that makes the action of quartic quasitopological
gravity well defined is the sum of all the boundary terms introduced above.
That is $I_{b}=I_{b}^{(1)}+I_{b}^{(2)}+I_{b}^{(3)}+I_{b}^{(4)}$.

In general, $I_{G}+I_{b}$ is divergent when evaluated on solutions, as is
other thermodynamic quantities. Rather than eliminating these divergences by
incorporating a reference term in the spacetime through the use of
substraction method of Brown and York \cite{subtrac}, one may add a new term
$I_{ct}$, which is a functional of the boundary curvature invariants \cite%
{Sken}. Since the boundary which we are interested in is flat, the proper
counterterm is
\begin{equation}
I_{ct}=-\frac{1}{8\pi }\int_{\partial \mathcal{M}}d^{n}x\sqrt{-\gamma }\;%
\frac{(n-1)}{L_{eff}},  \label{Ict}
\end{equation}%
where $L_{eff}$ is a scale length factor that depends on $l$ and the
coupling constants of gravity and should reduce to $l$ in the absence of
higher curvature terms.

\section{Charged black Brane Solutions\label{Solutions}}

Now, we introduce the charged black brane solutions of quasitopological
gravity in the presence of the nonlinear Maxwell field with Lagrangian (\ref%
{Lmat}). Using the static metric (\ref{met1}) with $g(\rho )=N^{2}(\rho
)f(\rho )$ and
\begin{equation}
A_{\mu }=h(\rho )\delta _{\mu }^{0}
\end{equation}%
for the vector potential, one can calculate the one dimensional action after
integration by parts. One obtains the action per unit volume as
\begin{equation}
I_{G}=\frac{{(n-1)}}{16\pi l^{2}}\int dtd\rho \lbrack {{{N(\rho )\left[ \rho
^{n}(1+\psi +\hat{\mu}_{2}\psi ^{2}+\hat{\mu}_{3}\psi ^{3}+\hat{\mu}_{4}\psi
^{4})\right] ^{\prime }+\frac{2^{s}l^{2}\rho ^{(n-1)}h^{\prime 2s}}{%
(n-1)N(\rho )^{2s-1}}}}}],  \label{Act3}
\end{equation}%
where $\psi =-l^{2}\rho ^{-2}f(\rho )$ and the dimensionless parameters $%
\hat{\mu}_{2}$, $\hat{\mu}_{3}$ and $\hat{\mu}_{4}$ are defined as:
\begin{equation*}
\hat{\mu}_{2}\equiv \frac{(n-2)(n-3)}{l^{2}}\mu _{2},\text{ \ \ \ }\hat{\mu}%
_{3}\equiv \frac{(n-2)(n-5)(3n^{2}-9n+4)}{8(2n-1)l^{4}}\mu _{3},
\end{equation*}%
\begin{equation*}
\hat{\mu}_{4}\equiv {\frac{n\left( n-1\right) \left( n-2\right) ^{2}\left(
n-3\right) \left( n-7\right) ({{n}^{5}-15\,{n}^{4}+72\,{n}^{3}-156\,{n}%
^{2}+150\,n-42)}}{{l}^{6}}}\mu _{4},
\end{equation*}%
Varying the action (\ref{Act3}) with respect to $\psi (r)$ yields
\begin{equation}
\left( 1+2\hat{\mu}_{2}\psi +3\hat{\mu}_{3}\psi ^{2}+4\hat{\mu}_{4}\psi
^{3}\right) \frac{dN(\rho )}{d\rho }=0,  \label{eom1}
\end{equation}%
which shows that $N(\rho )$ should be a constant which we set equal to one.
Variation with respect to $h(\rho )$ with $N(\rho )=1$ gives
\begin{equation}
(2s-1)\rho h^{\prime \prime }+(n-1)h^{\prime }=0,  \label{eom2}
\end{equation}%
and therefore
\begin{equation}
h(\rho )=\left\{
\begin{array}{cc}
\frac{q}{2}\ln (\rho ), & s=n/2 \\
-q\rho ^{-(n-2s)/(2s-1)}, & s\neq n/2%
\end{array}%
\right.  \label{hr}
\end{equation}%
where $q$ is an integration constant which is related to the charge
parameter. We pause to give some comments on the allowed values of $s$.
Equation (\ref{eom2}) shows that the electromagnetic field equation is
trivial for $s=1/2$. Also, in order to have a finite potential at infinity
and therefore a finite energy for the charged solutions, $s$ should be in the
range $1/2<s<n/2$. Thus in the rest of the paper, we consider only the
solutions for $1/2<s<n/2$.

Varying the action with respect to $N(\rho )$ and substituting $N(\rho )=1$
gives
\begin{equation}
\hat{\mu}_{4}\psi ^{4}+\hat{\mu}_{3}\psi ^{3}+\hat{\mu}_{2}\psi ^{2}+\psi
+\kappa =0,  \label{Eq4}
\end{equation}%
where%
\begin{equation*}
\kappa =1-\frac{m}{\rho ^{n}}+\frac{2^{s}l^{2}q^{2s}(n-2s)^{2s-1}}{%
(n-1)(2s-1)^{(2s-2)}\rho ^{2s(n-1)/(2s-1)}},
\end{equation*}%
and $m$ is an integration constant which is related to the mass of the
spacetime. The black brane solutions of (\ref{Eq4}), which reduces to black
brane solution in Einstein gravity in the absence of higher curvature terms
(the Einsteinian branch) is
\begin{equation}
f(\rho )=\frac{\rho ^{2}}{l^{2}}\left( \frac{\hat{\mu}_{3}}{4\hat{\mu}_{4}}+%
\frac{1}{2}R-\frac{1}{2}E\right) .  \label{F4}
\end{equation}%
where
\begin{eqnarray}
R &=&\left( \frac{{\hat{\mu}_{3}}^{2}}{4{\hat{\mu}_{4}}^{2}}-\frac{2\hat{\mu}%
_{2}}{3\hat{\mu}_{4}}+\left( {\frac{D}{2}+\sqrt{\Delta }}\right)
^{1/3}+\left( {\frac{D}{2}-\sqrt{\Delta }}\right) ^{1/3}\right) ^{1/2},
\label{RR} \\
E &=&\left( \frac{3{\hat{\mu}_{3}}^{2}}{4{\hat{\mu}_{4}}^{2}}-\frac{2\hat{\mu%
}_{2}}{\hat{\mu}_{4}}-R^{2}-\frac{1}{4R}\left[ \frac{4\hat{\mu}_{2}\hat{\mu}%
_{3}}{{\hat{\mu}_{4}}^{2}}-\frac{8}{\hat{\mu}_{4}}-\frac{{\hat{\mu}_{3}}^{3}%
}{{\hat{\mu}_{4}}^{3}}\right] \right) ^{1/2},  \label{EE} \\
\Delta &=&\frac{H^{3}}{27}+\frac{D^{2}}{4},\text{ \ \ \ \ \ \ }H={\frac{3%
\hat{\mu}_{3}-{\hat{\mu}_{2}}^{2}}{3{\hat{\mu}_{4}}^{2}}}-\,{\frac{4\kappa }{%
\hat{\mu}_{4}},} \\
D &=&{\frac{2}{27}}\,{\frac{{\hat{\mu}_{2}}^{3}}{{\hat{\mu}_{4}}^{3}}}-\frac{%
1}{3}\,\left( {\frac{\hat{\mu}_{3}}{{\hat{\mu}_{4}}^{2}}}+8\,{\frac{\kappa }{%
\hat{\mu}_{4}}}\right) \frac{\hat{\mu}_{2}}{\hat{\mu}_{4}}+{\frac{{\hat{\mu}%
_{3}}^{2}\kappa }{{\hat{\mu}_{4}}^{3}}}+\frac{1}{{\hat{\mu}_{4}}^{2}}.
\end{eqnarray}

The metric function $f(\rho )$ for the uncharged solution $(q=0)$ is real in
the whole range $0\leq \rho <\infty $. But for charged solutions, one should
restrict the spacetime to the region $\rho \geq r_{0}$, where $r_{0}$ is the
largest real root of
\begin{eqnarray}
&&{\frac{1}{4}}\,\left[ -{\frac{2}{27}}\,{\frac{{\hat{\mu}}_{{2}}^{3}}{{\hat{%
\mu}}_{{4}}^{3}}}+\frac{1}{{3\hat{\mu}_{{4}}}}\,\left( {\frac{\hat{\mu}_{{3}}%
}{{\hat{\mu}}_{{4}}^{2}}}-\,{\frac{4\kappa _{0}}{\hat{\mu}_{{4}}}}\right)
\hat{\mu}_{{2}}-{\frac{{\hat{\mu}_{{3}}}^{2}\kappa _{0}}{{\hat{\mu}}_{{4}%
}^{3}}}+{\frac{4\hat{\mu}_{{2}}\kappa _{0}}{{\hat{\mu}_{{4}}}^{2}}}-\frac{1}{%
{\hat{\mu}_{{4}}}^{2}}\right] ^{2}  \notag \\
&&\hspace{5cm}+{\frac{1}{27}}\,\left[ {\frac{\hat{\mu}_{{3}}}{{\hat{\mu}_{{4}%
}}^{2}}}-\,{\frac{4\kappa _{0}}{\hat{\mu}_{{4}}}}-\,{\frac{{\hat{\mu}_{{2}}}%
^{2}}{3{\hat{\mu}_{{4}}}^{2}}}\right] ^{3}=0
\end{eqnarray}%
where
\begin{equation}
\kappa _{0}=1-\frac{m}{r_{0}^{n}}+\frac{2^{s}l^{2}q^{2s}(n-2s)^{2s-1}}{%
(n-1)(2s-1)^{(2s-2)}r_{0}^{2s(n-1)/(2s-1)}},
\end{equation}

Performing the transformation
\begin{equation}
r=\sqrt{{\rho }^{2}-{r_{0}}^{2}}\Rightarrow d{\rho }^{2}=\frac{r^{2}}{%
r^{2}+r_{0}^{2}}dr^{2}
\end{equation}%
the metric becomes%
\begin{equation}
ds^{2}=-f(r)dt^{2}+\frac{{r^{2}dr^{2}}}{(r^{2}+r_{0}^{2})f(r)}%
+(r^{2}+r_{0}^{2})\;\sum_{i=1}^{n-1}d\phi _{i}{}^{2}.  \label{metr0}
\end{equation}%
where now the functions $h(r)$ and $\kappa $ are
\begin{eqnarray}
h(r) &=&-q(r^{2}+r_{0}^{2})^{(2s-n)/(4s-2)},  \label{hr2} \\
\kappa &=&1-\frac{m}{(r^{2}+r_{0}^{2})^{n/2}}+\frac{%
2^{s}l^{2}q^{2s}(n-2s)^{2s-1}}{%
(n-1)(2s-1)^{(2s-2)}(r^{2}+r_{0}^{2})^{s(n-1)/(2s-1)}},
\end{eqnarray}%
The geometrical mass of black brane solutions in terms of the horizon
radius, $r_{+}$, can be obtained as
\begin{equation}
m=\left\{ 1+\frac{%
2^{s}l^{2}q^{2s}(n-2s)^{2s-1}(r_{+}^{2}+r_{0}^{2})^{s(1-n)/(2s-1)}}{%
(n-1)(2s-1)^{(2s-2)}}\right\} (r_{+}^{2}+r_{0}^{2})^{n/2},
\end{equation}

\section{Thermodynamics of Charged Black Branes \label{Therm}}

The Hawking temperature for the black brane solution given in \ the last
section can be calculated as:
\begin{equation}
T=\frac{f^{\prime }(r_{+})}{4\pi }\sqrt{1+\frac{r_{0}^{2}}{r_{+}^{2}}}=\{%
\frac{n}{4\pi {l}^{2}}-\frac{q^{2s}2^{s}{(n-2s)}^{2s}(r_{+}^{2}+r_{0}^{2})^{{%
s(1-n)}/{(2s-1)}}}{4\pi (n-1)(2s-1)^{2s-1}}\}\sqrt{r_{+}^{2}+r_{0}^{2}},
\label{Temp0}
\end{equation}%
The electric potential $\Phi $, measured at infinity with respect to the
horizon, is defined by \cite{Gub}
\begin{equation}
\Phi =A_{\mu }\xi ^{\mu }\mid _{r\rightarrow \infty }-A_{\mu }\xi ^{\mu
}\mid _{r=r_{+}},  \label{Pot}
\end{equation}%
where $\xi ^{\mu }$ is the null generator of the horizon, $A_{\mu }$ is just
the vector potentia.One obtains
\begin{equation}
\Phi =\frac{q}{(r_{+}^{2}+r_{0}^{2})^{(n-2s)/2(2s-1)}},  \label{Ph0}
\end{equation}%
Using Gibbs free energy, we can calculate the entropy, electric charge and
mass of this black hole. The Gibbs free energy in term of the on-shell
action is $G=TI_{E}$, where $I_{E}$ is the Euclidean action and $T$ is the
temperature \cite{Gib}.

The finite total action, $I_{E}=I_{g}+I_{b}+I_{ct}$ is finite, if $L_{eff}$
\ of Eq. (\ref{Ict}) is chosen as
\begin{equation}
L_{eff}=-\frac{210\psi _{\infty }^{1/2}}{15\hat{\mu _{4}}\psi _{\infty
}^{4}+21\hat{\mu _{3}}\psi _{\infty }^{3}+35\hat{\mu _{2}}\psi _{\infty
}^{2}-105\psi _{\infty }-105}l,  \label{leff}
\end{equation}%
where $\psi _{\infty }$ is the limit of $\psi $ at infinity for our solution
given in Eq. (\ref{F4}). The finite action per unit volume can be calculated
as
\begin{equation}
I_{E}=-\frac{\beta }{16\pi l^{2}}\left\{ (r_{+}^{2}+r_{0}^{2})^{n/2}+\frac{%
2^{s}l^{2}q^{2s}{(n-2s)}^{2s-1}(r_{+}^{2}+r_{0}^{2})^{{(2s-n)}/{2(2s-1)}}}{{%
(2s-1)}^{2s-2}(n-1)}\right\} ,  \label{FAction}
\end{equation}

Having the total finite action, one can obtain the Gibbs free energy per
unit volume as
\begin{equation}
G(T,\Phi )=-\frac{1}{16\pi l^{2}}\left\{ (r_{+}^{2}+r_{0}^{2})^{n/2}+\frac{%
2^{s}l^{2}{(n-2s)}^{2s-1}\Phi ^{2s}(r_{+}^{2}+r_{0}^{2})^{{(n-2s)}/{2}}}{{%
(2s-1)}^{2s-2}(n-1)}\right\} ,  \label{Gib1}
\end{equation}%
where the implicit dependence of $r_{+}^{2}+r_{0}^{2}\equiv \rho {_{+}}^{2}$
in terms of the intensive quantities $T$ and $\Phi $ is%
\begin{equation*}
n(n-1)\rho {_{+}^{2}}-4\pi l^{2}(n-1)T\rho {_{+}}-2^{s}\left(n-2s\right) ^{2s}(2s-1)^{1-2s}l^{2}{\Phi }^{2s}\rho {_{+}^{2(1-s)}}=0.
\end{equation*}%
It is notable that for $s=1$, the above equation may be solved for $\rho {%
_{+}}$, and therefore one can find the explicit form of $\rho {_{+}}$ in
terms of $T$ and $\Phi $ \cite{DV}. Here, one can calculate $(\partial T/\partial \rho _{+})_{\Phi}$
and $(\partial \Phi/\partial \rho _{+})_{T}$ and
the entropy and charge per unit volume of charged black brane may be
obtained by using the well-known equations
\begin{eqnarray}
S &=&-\left( \frac{\partial G}{\partial T}\right) _{\Phi }=-\left( \frac{%
\partial G}{\partial \rho _{+}}\right) _{\Phi }\left( \frac{\partial T}{\partial \rho
_{+}}\right)^{-1}_{\Phi }=\frac{1}{4}(r_{+}^{2}+r_{0}^{2})^{(n-1)/2},
\label{Ent1} \\
Q &=&-\left( \frac{\partial G}{\partial \Phi }\right) _{T}=-\left( \frac{%
\partial G}{\partial \Phi }\right) _{T}-\left( \frac{\partial G}{\partial
\rho _{+}}\right)_{T}\left( \frac{\partial \Phi}{\partial \rho _{+} }%
\right)^{-1}_{T}=\frac{2^{s}s{(n-2s)^{2s-1}}q^{2s-1}}{8\pi {(2s-1)}^{2s-1}},
\label{Charge1}
\end{eqnarray}%
which show that they are in agreement with the entropy density calculated in
Ref. \cite{DBM} and the charge density that may be obtained by use of Gauss
theorem, respectively.

In order to check the first law of thermodynamics of the black brane, one
may write the energy density as a function of extensive parameters, $S$ and $%
Q$ as
\begin{eqnarray}
M(S,Q) &=&G+TS+\Phi Q=\frac{n-1}{16\pi }m \\
&=&\frac{n-1}{16\pi }\left[ (4S)^{2s/(2s-1)}+\frac{(8\pi
/s)^{2s/(2s-1)}l^{2}(2s-1)^{2}}{2^{s/(2s-1)}(n-1)(n-2s)}Q^{2s/(2s-1)}\right]
(4S)^{(2s-n)/(n-1)}.  \notag
\end{eqnarray}%
The first law of thermodynamics, $dM=TdS+\Phi dQ$, is satisfied. This is due
to the fact that the intensive quantities
\begin{equation}
T=\left( \frac{\partial M}{\partial S}\right) _{Q},\,\ \ \ \ \Phi =\left(
\frac{\partial M}{\partial Q}\right) _{S}.  \label{Inten}
\end{equation}%
are in agreement with those given in Eqs. (\ref{Temp0}) and (\ref{Ph0}).

Also, it is worth to mention that the energy per unit volume of the black
brane in terms of the thermodynamics quantities can be written as%
\begin{equation*}
M=\frac{n-1}{n}\left( TS+Q\Phi \right) .
\end{equation*}

\section{Thermodynamics of Charged Rotating Black Branes \label{Rot}}

Here we consider the metric of ($n+1$)-dimensional rotating spacetime with
flat horizon and $k$ rotation parameters, which may be written as \cite%
{{Awad:2002cz}}
\begin{eqnarray}
ds^{2} &=&-f(r)\left( \Xi dt-{{\sum_{i=1}^{k}}}a_{i}d\phi _{i}\right) ^{2}+%
\frac{r^{2}+r_{0}^{2}}{l^{4}}{{\sum_{i=1}^{k}}}\left( a_{i}dt-\Xi l^{2}d\phi
_{i}\right) ^{2}  \notag \\
&&\ +\frac{{r^{2}dr^{2}}}{(r^{2}+r_{0}^{2})f(r)}-\frac{r^{2}+r_{0}^{2}}{l^{2}%
}{\sum_{i<j}^{k}}(a_{i}d\phi _{j}-a_{j}d\phi _{i}\phi
_{i})^{2}+(r^{2}+r_{0}^{2})\;\sum_{i=k+1}^{n-1}d\phi _{i}{}^{2},
\label{met2}
\end{eqnarray}%
where the angular coordinates are in the range $0\leq \phi _{i}<2\pi $, $\Xi
=\sqrt{1+\mathbf{a}^{2}/l^{2}}$ with $\mathbf{a}^{2}={{\sum_{i=1}^{k}a}}%
_{i}^{2}$and $f(r)$ is the same as the metric function in static case. One
should note that the maximum number of rotation parameter in $(n+1)$
dimensions is $k=[n/2]$, where $[x]$ is the integer part of $x$. Although
the metrics (\ref{metr0}) and (\ref{met2}) can be mapped into each other
locally by the transformation%
\begin{equation*}
t^{\prime }=\Xi t-\sum_{i=1}^{k}a_{i}\phi _{i},\,\ \ \ \ \ \phi _{i}^{\prime
}=\frac{a_{i}}{l^{2}}t-\Xi \phi _{i},
\end{equation*}%
they cannot mapped into each other globally. This is due to the periodic
nature of the coordinates $\phi _{i}$, $i=1...k$ \cite{Stach}.

Now, we investigate the thermodynamics of rotating charged black branes. The
gauge potential for this solution is given by
\begin{equation}
A_{\mu }=h(r)\left( \Xi dt-\sum_{i=1}^{k}a_{i}d\phi _{i}\right) ,
\end{equation}%
where $h(r)$ is given in Eq. (\ref{hr2}). The temperature of the Killing
horizon is given by
\begin{equation}
T=\frac{1}{2\pi }\left( -\frac{1}{2}\nabla _{b}\xi _{a}\nabla ^{b}\xi
^{a}\right) _{r=r_{+}}^{1/2},  \label{Tem1}
\end{equation}%
where $\xi $ is the combination of Killing vectors
\begin{equation}
\xi =\partial _{t}+\sum_{i=1}^{k}\Omega _{i}\partial _{\phi _{i}}
\label{xii}
\end{equation}%
and $\Omega _{i}$ is the angular velocity of the horizon given as
\begin{equation}
\Omega _{i}=-\left[ \frac{g_{t\phi _{i}}}{g_{\phi _{i}\phi _{i}}}\right]
_{r=r_{+}}=\frac{a_{i}}{\Xi l^{2}}.  \label{Om}
\end{equation}%
Using Eqs. (\ref{Tem1})-(\ref{Om}), one obtains
\begin{equation}
T=\frac{1}{\Xi }\left\{ \frac{n}{4\pi {l}^{2}}-\frac{q^{2s}2^{s}(2s-1)[{%
(n-2s)}/{(2s-1)}]^{2s}(r_{+}^{2}+r_{0}^{2})^{{s(1-n)}/{(2s-1)}}}{4\pi (n-1)}%
\right\} \sqrt{r_{+}^{2}+r_{0}^{2}}.  \label{Temp1}
\end{equation}

The electric potential $\Phi $ measured at infinity with respect to the
horizon is
\begin{equation}
\Phi =\frac{1}{\Xi }\frac{q}{(r_{+}^{2}+r_{0}^{2})^{(n-2s)/2(2s-1)}}.
\label{phi1}
\end{equation}%
$\mathbf{\ }$The finite action per unit volume can be calculated through the
use of the counterterm method as
\begin{equation}
I_{E}=-\frac{\beta }{16\pi l^{2}}\left\{ (r_{+}^{2}+r_{0}^{2})^{n/2}+\frac{%
2^{s}l^{2}q^{2s}{(n-2s)}^{2s-1}(r_{+}^{2}+r_{0}^{2})^{{(2s-n)}/{2(2s-1)}}}{{%
(2s-1)}^{2s-2}(n-1)(1-l^{2}\mathbf{\Omega }^{2})}\right\} .  \label{Actrot}
\end{equation}

Using Eqs. (\ref{Om}), (\ref{Temp1}) and (\ref{phi1}), the Gibbs free energy
per unit volume in terms of the intensive quantities $T$, $\Phi $ and $%
\Omega _{i}$'s may be written as
\begin{equation}
G(T,\Phi ,\mathbf{\Omega })=-\frac{1}{16\pi l^{2}}\left\{
(r_{+}^{2}+r_{0}^{2})^{n/2}+\frac{2^{s}l^{2}{(n-2s)}^{2s-1}\Phi
^{2s}(r_{+}^{2}+r_{0}^{2})^{{(n-2s)}/{2}}}{{(1-l^{2}\mathbf{\Omega }%
^{2})(2s-1)}^{2s-2}(n-1)}\right\} ,  \label{Gibrot}
\end{equation}%
where $\mathbf{\Omega }^{2}={{\sum_{i=1}^{k}\Omega }}_{i}^{2}$. Now using
the well-known formulas of entropy, charge, energy per volume of rotating
charged black brane in term of the Gibbs free energy and the relation
between $\eta _{+}$, $T$ and $S$, one obtains
\begin{equation}
S=-\left( \frac{\partial G}{\partial T}\right) _{\Phi ,\Omega _{i}}=\frac{%
\Xi }{4}(r_{+}^{2}+r_{0}^{2})^{(n-1)/2},  \label{Ent2}
\end{equation}%
\begin{equation}
Q=-\left( \frac{\partial G}{\partial \Phi }\right) _{T,\Omega _{i}}=\Xi
\frac{\lbrack {(n-2s)/(2s-1)}]^{2s-1}2^{s}sq^{2s-1}}{8\pi }.
\end{equation}%
\begin{equation}
J_{i}=-\left( \frac{\partial G}{\partial \Omega _{i}}\right) _{T,\Phi }=%
\frac{n}{16\pi }\Xi ma_{i}.
\end{equation}%
The energy density per unit volume may be written as%
\begin{equation*}
M=G+TS+\Phi Q+\sum\limits_{i=1}^{k}\Omega _{i}J_{i}=\frac{1}{16\pi }(n\Xi
^{2}-1)m.
\end{equation*}%
Again, one may confirm the first law of thermodynamics by computing the
angular velocities, temperature and potential
\begin{equation}
\text{\ }\Omega _{i}=\left( \frac{\partial M}{\partial J_{i}}\right) _{S,Q},%
\text{ \ \ }T=\left( \frac{\partial M}{\partial S}\right) _{Q,\mathbf{J}%
},\,\ \ \ \ \Phi =\left( \frac{\partial M}{\partial Q}\right) _{S,\mathbf{J}%
},
\end{equation}%
and compare them with $\Omega _{i},$ $T$ and $\Phi $ given in Eqs. (\ref{Om}%
), (\ref{Temp1}) and (\ref{phi1}), respectively.

\section{Concluding Remarks \label{Conc}}

In this paper, our first aim was to introduce a surface term that makes the
action of quartic quasitopological gravity well defined. This was done by
introducing seven terms of order seven in extrinsic curvature of the
boundary with suitable coefficients in such a way that cancel the derivative
of the variation of the metric normal to the flat boundary. Although, we
have used this boundary term in order to calculate the on-shell action of
black hole solution of the quasitopological gravity, one may use it in
Hamiltonian formalism of quasitopological gravity. Having the nonlinear
terms in Riemann tensor in the action of gravity, we assumed to have
nonlinear terms in the invariant quantity of the electromagnetic field
tensor too. That is, we considered the quartic quasitopological gravity in
the presence of nonlinear electromagnetic field introduced in \cite{Martinez}%
, and introduced the black brane solutions of this theory. These solutions
may be interpreted as black brane solutions with two inner and outer event
horizons or an extreme black brane depending on the value of charge
parameter.

In order to investigate the thermodynamics of these charged solutions, one
needs the finite on-shell action of the solutions. But, as in the case of
Einstein solutions, the action and the conserved quantities of the solutions
of quasitopological gravity are not finite. We used the counterterm method
to calculate the finite action. Using the definition of Gibbs free energy
for the solutions, we computed the thermodynamic quantities $S$, $Q$ and the
energy of the solutions per unit volume. We found that the thermodynamic and
conserved quantities of black branes of quasitopological gravity with flat
horizon are independent of the the Gauss-Bonnet and quasi topological
coefficients (${\mu _{2}}$, ${\mu _{3}}$ and ${\mu _{4}}$). We obtained the
mass as a function of the extensive parameters $S$ and $Q$ and showed that
the conserved and thermodynamic quantities satisfy the first law of
thermodynamic. Finally we generalized the solutions to the rotating
solutions and obtained the conserved and thermodynamic quantities of these
rotating charged solutions. The surface terms of quartic quasitopological
gravity introduced here may be used in the investigation of the
thermodynamics of asymptotic Lifshitz black branes, which will be given
elsewhere.

\acknowledgements This work has been supported financially by Center for
Excellence in Astronomy \& Astrophysics (CEAA -- RIAAM) under research project No. 1/3232.

\end{document}